\documentclass[dvips]{acta}
\usepackage{supertabular,lscape,epsfig}
\usepackage{graphicx}
\usepackage{amssymb}
\usepackage{amsmath}
\DeclareMathAlphabet{\mathitbf}{OML}{cmm}{b}{it}

\SetPages{0}{0}

\SetVol{62}{2011}

\begin{document}

\begin{Titlepage}
\Title{Open clusters in 2MASS photometry -- I. Structural and basic
astrophysical parameters}

\Author{{\L}. ~B~u~k~o~w~i~e~c~k~i, ~G. ~M~a~c~i~e~j~e~w~s~k~i, ~P.
~K~o~n~o~r~s~k~i ~and ~A. ~S~t~r~o~b~e~l}{Toru\'n Centre for Astronomy,
Nicolaus Copernicus University, Gagarina 11, PL-87-100 Toru\'n, Poland\\
e-mail: bukowiecki@astri.umk.pl}

\Received{June 5, 2011}
\end{Titlepage}

\Abstract{The main goal of our
project is to obtain a complete picture of individual open clusters from
homogeneous data and then search for correlations between their
astrophysical parameters. The near--infrared $JHK_{S}$ photometric data
from the 2--Micron All Sky Survey were used to determine new coordinates
of the centres, angular sizes and radial density profiles for 849 open
clusters in the Milky Way. Additionally, age, reddening, distance, and
linear sizes were also derived for 754 of them. For \textbf{these} open clusters
our results are in satisfactory agreement with the literature data. The
analysed sample contains open clusters with ages in the range from 7 Myr
to 10 Gyr. The majority of these clusters are located up to 3 kpc from
the Sun, less than 0.4 kpc from the Galactic Plane and 6 -- 12 kpc from
the Galactic Centre. The majority of clusters have core radii of about
1.5 pc and the limiting radii of the order of 10 pc. We notice that in
the near--infrared, open clusters seem to be greater than in optical
bands. We notice that a paucity of clusters is observed at Galactic
longitudes range from $140\,^{\circ}$ to $200\,^{\circ}$ which probably
reflects the real spatial distribution of open
clusters in the Galaxy. The lack of clusters was also found in
earlier studies.}

{open clusters and associations: general
-- infrared: galaxies -- astronomical databases: 2MASS}


\section{Introduction}

In this paper the near-infrared $JHK_{S}$ photometric data were used to
redetermine equatorial and galactic coordinates of the centres of known
open clusters. This allowed us to construct radial density profiles of
clusters and to determine their core radii and estimate their limiting
radii. The adopted background-star decontaminating procedure allowed us
to construct colour--magnitude diagrams for individual clusters and to
estimate ages for a significant fraction of the considered sample.

 The structure of this paper is as follows. In
Section 2 the method and data analysis is presented. In Section 3 our results are
compared to the results found in literature. Section 4 contains
discussion of the relations between individual  parameters. Final
conclusions are summarized in Section 5.

\textbf{Similar to our research, Janes (1979) used $\mathitbf{UBV}$ photometry 
to study the reddening and metallicity of 41 open clusters. Comparison between the age and position in the Galaxy was studied by Lyng\aa~(1980, 1982). Janes \& Phelps (1994) based on 72 open clusters and Friel (1995) with \textbf{a} sample \textbf{of} 74 objects \textbf{investigated} how the extinction and age depend on \textbf{a} position in the Galaxy. Tadross et al.~(2002) and Schilbach et al.~(2006) studied how open clusters diameters and age depend on position in the Galactic Plane. Previous research of open clusters in 2MASS photometry were done by Dutra \& Bica (2001) who studied 42 new infrared star clusters, stellar groups and candidates towards the Cyngus X region. Bica et al.~(2003) based on 2MASS \textbf{researched} 346 open clusters, whereas 315 objects were previously known. They studied linear diameters and spatial distribution of open clusters in the Galaxy. Recently Froebrich et al.~(2010) studied 269 open clusters and the age, core radius, the reddening, Galactocentric distance and the scaleheight were determined.}


\section{Data analysis}

\subsection{Data source and selection of clusters}

The list of open clusters and their initial central coordinates were
taken from the compilation \emph{New catalog of optically visible open
clusters and candidates} by Dias et al.~(2002). Our research is based on
the $JHK_{S}$ photometric data extracted from the
2MASS\footnote{http://www.ipac.caltech.edu/2mass/releases/allsky/}
\emph{Point Source Catalog} (Skrutskie et al.~ 2006). The adopted
extraction radius for each listed cluster was equal to 30 + 4 arcmin,
because we wanted to focus on well defined, concentrated open clusters 
of angular diameters smaller than 60 arcmin. The
additional 4 arcmin were added as a margin for cases in which the true
centres were found at different location. This step resulted in
a collection of 1756 catalogue fields.

The next step was to identify the star cluster in each of extracted field.
In many cases this task was not trivial because some clusters were found
to be poor, sometimes consisting of only from a few  to several dozen
members, and sparse due to a significant angular size. Moreover, the
stellar background in near IR is relatively high, making the
identification of a
cluster very uncertain or even impossible. In some cases, it
was necessary to exclude open clusters from further consideration as
they were too wide or not rich enough for  reliable analysis.
Finally, after a visual inspection of pictures from the 2MASS and
iterative construction of radial density profiles (RDP) for 1756
clusters, we limited our sample to 849 unambiguously identified objects
(see Sect. 2.2 and 2.3) for which reliable centre coordinates could be
determined.

\subsection{Central coordinates}

Visual inspection of the extracted fields suggested that for the
majority of open clusters their catalogue coordinates were much
different from the real ones. Hence, they must have been redetermined.
The algorithm for determination of the cluster centre was adopted from
Maciejewski and Niedzielski (2007). The procedure starts with initial
coordinates of cluster centres taken from the catalogue by Dias et al.~
(2002). In some cases, the discrepancies were so significant that it was
necessary to define the initial coordinates by hand as a result of
visual inspection. In the considered field, the algorithm cuts off two
perpendicular rectangular stripes along declination and right ascension,
crossed over an approximated cluster centre, and a histogram of star
counts is built along each stripe. The length and width of those stripes
are chosen individually for each cluster and depend on its size.
Typically, the length ranges from 5 to 30 arcmin and the width from 1 to
8 arcmin. Bins with the positions of the maximum value in both
coordinates indicate a new cluster central coordinates. This procedure
was repeated iteratively until the new position of the cluster centre
converged into a stable point on the sky.

\subsection{Angular sizes}

To study the cluster structure, a radial density profile was constructed
for each cluster by  star counts inside concentric
rings centred at the redetermined centre coordinates.
The observed stellar density $\rho$ was calculated for each
ring and plotted as a function of the angular radial distance from the
cluster centre. Then, this density distribution was parametrised by a
two-parameter King-like function (King 1966):  \begin{equation}
 \rho(r)=f_{bg}+\frac{f_{0}}{1+\left(\frac{r}{r_{core}}\right)^{2}} \, , \;
\end{equation}
where $r_{core}$, $f_{0}$, and $f_{bg}$ are the core radius, the central
density, and the background density level, respectively. The core radius
was defined as a distance where the stellar density drops to half of
$f_{0}$. These parameters were determined by the least-square method. The
cluster limiting radius, $r_{lim}$, was calculated by comparing
$\rho(r)$ to a border background density level, $\rho_{b}$, defined as:
\begin{equation}
 \rho_{b}=f_{bg}+3\sigma_{bg} \, ,  \;
\end{equation}
where $\sigma_{bg}$ is uncertainty of $f_{bg}$. Finally
, $r_{lim}$ was calculated according to the following formula: 
\begin{equation}
r_{lim}=r_{core}\sqrt{\frac{f_{0}}{3\sigma_{bg}}-1} \, . \;
\end{equation}

The widths of concentric rings were set equal to uncertainties (errors) of
$r_{lim}$ and ranged from 0.14 to 4.4 arcmin,
depending on \textbf{the} angular diameter of each cluster and stellar crowding. They were
further refined individually for each object to obtain the RDP as smooth as
possible. We combined both $r_{core}$ and $r_{lim}$ in a form of the
concentration parameter $c=\log {(r_{lim}}/{r_{core})}$ (Peterson and
King 1975) to characterise the structure of clusters. The structural
parameters -- new equatorial and galactic coordinates and angular sizes
of all object in the sample -- are given in Table~1 in Appendix A.

\subsection{The colour--magnitude diagram analysis}

Analysis of colour--magnitude diagrams (CMDs) is a commonly used method
to derive age, reddening, and distance -- the most fundamental
parameters characterising stellar clusters. For each cluster of our
sample, a $J$ \textit{vs.} ($J-K_{S}$) colour--magnitude diagram was
constructed. Although in general it is impossible to point out individual
cluster members  without
additional information, \eg proper motion, the
contribution from field stars can be removed from the cluster CMD in a
statistical sense. The algorithm applied to our data
was based on the idea , presented in Mighell et al.~(1996) and discussed in
Bica and Bonatto (2005), of statistical subtracting a CMD of a
background field outside of the cluster area from a CMD of a field
occupied by a cluster. Because of various contrast of individual
clusters against the Galaxy background, the boundary of the area
occupied by a cluster was chosen individually for each cluster as a
result of iterative tests. In practice values between 1 and 4 $r_{core}$
were used, depending on a ratio $r_{lim}/r_{core}$ (based on our
calculations). When this ratio was greater than 9 we used 3-4
$r_{core}$, when was less than 5.5 we used 1-2 $r_{core}$, for other
cases we used 2-3 $r_{core}$ (exact values of $r_{core}$ are listed in
Table~1 in Appendix A). A concentric offset field began at $r = r_{lim}
+1$ arcmin from the cluster centre and usually ended at 34 arcmin. Both
CMDs were divided into two--dimensional bins of $\Delta J = 0.4$ mag and
$\Delta (J-K_{S})=0.1$ mag size (both values being fixed after a series
of tests, as a compromise between resolution and the star numbers in
individual boxes). The number of stars within each box was counted. The
cleaned (decontaminated) CMD was built by subtracting a number of stars
in an offset box from a number of stars in a corresponding cluster box.
The latter number was weighted by the cluster--to--offset field area
ratio. According to this procedure the algorithm randomly chose the
required number of stars located in the cluster area and with the
adequate magnitude and colour index. Finally, the list of stars in each
cleaned cluster box was saved and used for constructing the
decontaminated CMD.

After applying the CDM cleaning procedure, we limited our sample to 754
open clusters with well visible main sequence  and red giant clump
(excluding very young clusters). Stars which were found to be outliers
from main sequence and red giant clump were removed by hand in a result
of visual inspection. Finally, the distances moduli, reddenings, and
ages were derived by fitting the Padova isochrones at $J$ \textit{vs.}
$J-K_{S}$ bands for solar metallicity $Z = 0.019$ (Girardi et al.~ 2002)
to the decontaminated CMDs. The isochrones cover a wide range of ages
from 4 Myr up to 14 Gyr with a step of 0.05 in the logarithm of age. 
Theoretical isochrones were shifted in both directions in the CMDs
with a step of 0.01 mag. The solution giving the smallest $\chi^{2}$ was
taken as a final one. In a series of tests we checked how the reddening
and distance modulus determination depend on choosing different angular
sizes of both cluster and offset--field areas. We found that the spread
of results, represented by the standard deviation, usually did not
exceed 0.03 and 0.10 mag, for the reddening and distance modulus,
respectively. \textbf{Step of 0.05 in the logarithm of age} was adopted as a
typical uncertainty of the log($age$)
but for younger clusters for which we could not detect giant stars and
we used only ZAMS stars, the individual error can be greater. In these
cases our values of age are more like ``not older than''. \textbf{We present six
CDMs for open clusters with different age in Appendix B.}

Results from the analysis of CMDs allowed us to calculate cluster
distances from the Sun and transform angular sizes into linear ones. The
determined astrophysical fundamental parameters are listed in Table~2 in
Appendix A.


\section{Comparison with the published data}

To test a reliability of our determination, we compared our results with
the literature data. We took them from the open cluster catalogue by
Dias et al.~(2002) or from papers dedicated to individual clusters, if
available. Fig.~1a presents comparison of angular limiting radii
obtained by us in the near IR with data in visual bands from the
literature for 337 open clusters. Fig.~1b presents the same comparison
but for angular core radii and for 85 objects. As it can be easily seen,
for the majority of observed open clusters our calculations of angular
sizes (near IR) appear to be even several times larger than the
catalogue records indicate. This has already been noticed by Sharma et
al.~(2006). This effect seems to be particularly distinct for small --
according to catalogue data -- objects, for which our estimations of
this parameter appear to be up to six times greater. \textbf{This} effect may be a
result of missing faint cluster members in many photometric studies
\textbf{which are} based on optical photometry, partly due to significant influence of the
interstellar extinction or using the field of view not wide
enough. Using the 2MASS survey allows us to avoid both of these limits:
provide wide cluster fields coverage with homogenous data and minimise
the interstellar absorption in the infrared.

\begin{figure}[htb]
\begin{center}
\includegraphics[width=1.0\textwidth]{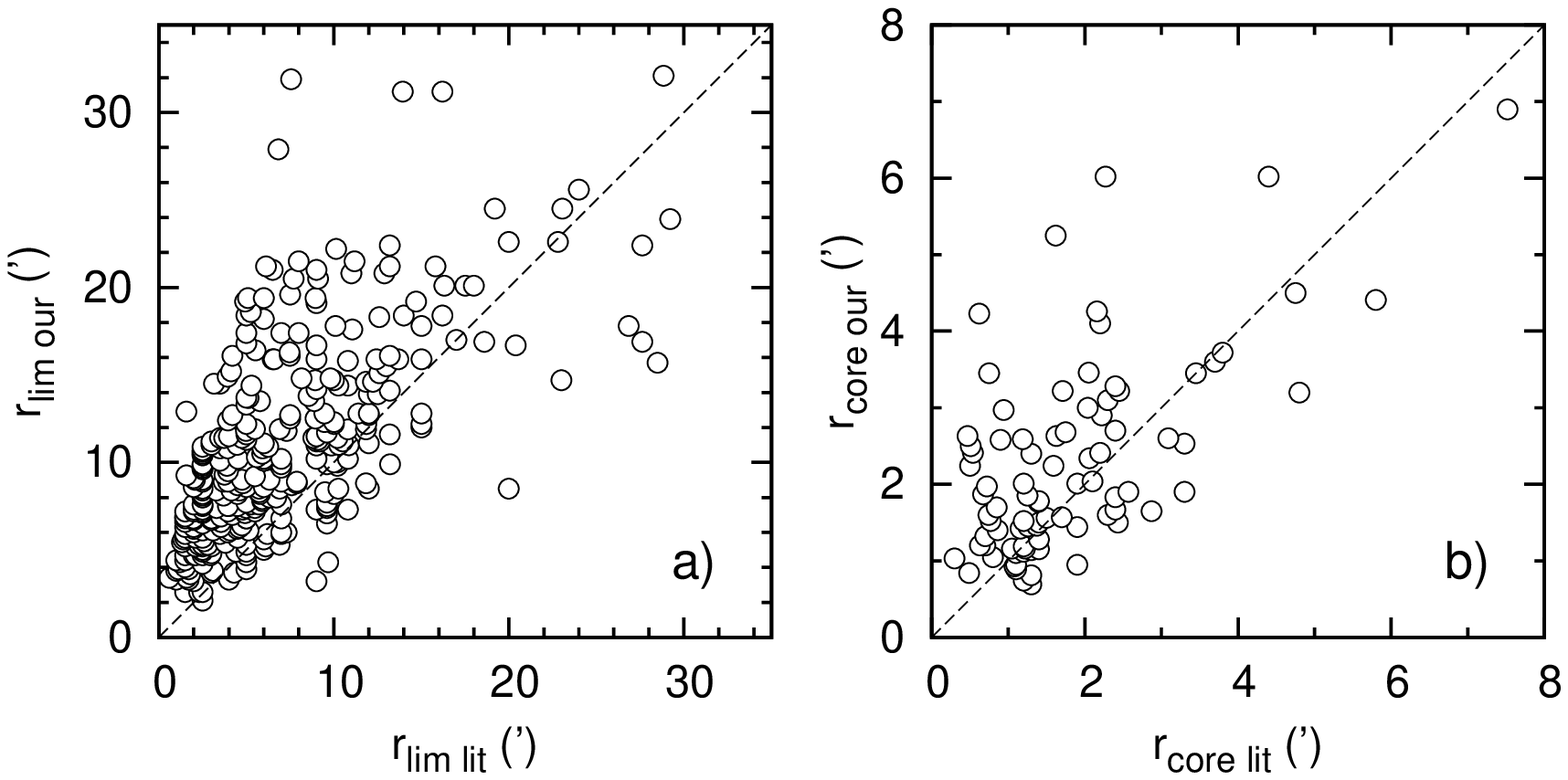}
\end{center}
\FigCap{a) A comparison of angular limiting radii obtained by us in
near-IR ($JHK$) and the literature data obtained in visual bands
($UBV$) for 337 open clusters. The dotted line shows a case of perfect
match, i.e. $y=x$. b) The same as panel a but for angular core radii for
85 clusters.} 
\end{figure}

We expected that ages of open clusters determined by us will be very close
to the literature ones because the use of data from different
photometric bands should have little influence on determining
this parameter. In our sample we found 536 open clusters with known
ages. In Fig.~2a our results are plotted against the published values.
The majority of points is located near a straight line corresponding to
a perfect match. Thirteen outlying points, marked as \textbf{triangles}, were
rejected during fitting a linear relation. We obtained:

\begin{equation}
log\:(age)_{our} = (1.05\pm0.06)\:log\:(age)_{lit}
\end{equation}

with the correlation coefficient of 0.96.

\begin{figure}[htb]
\begin{center}
\includegraphics[width=1.0\textwidth]{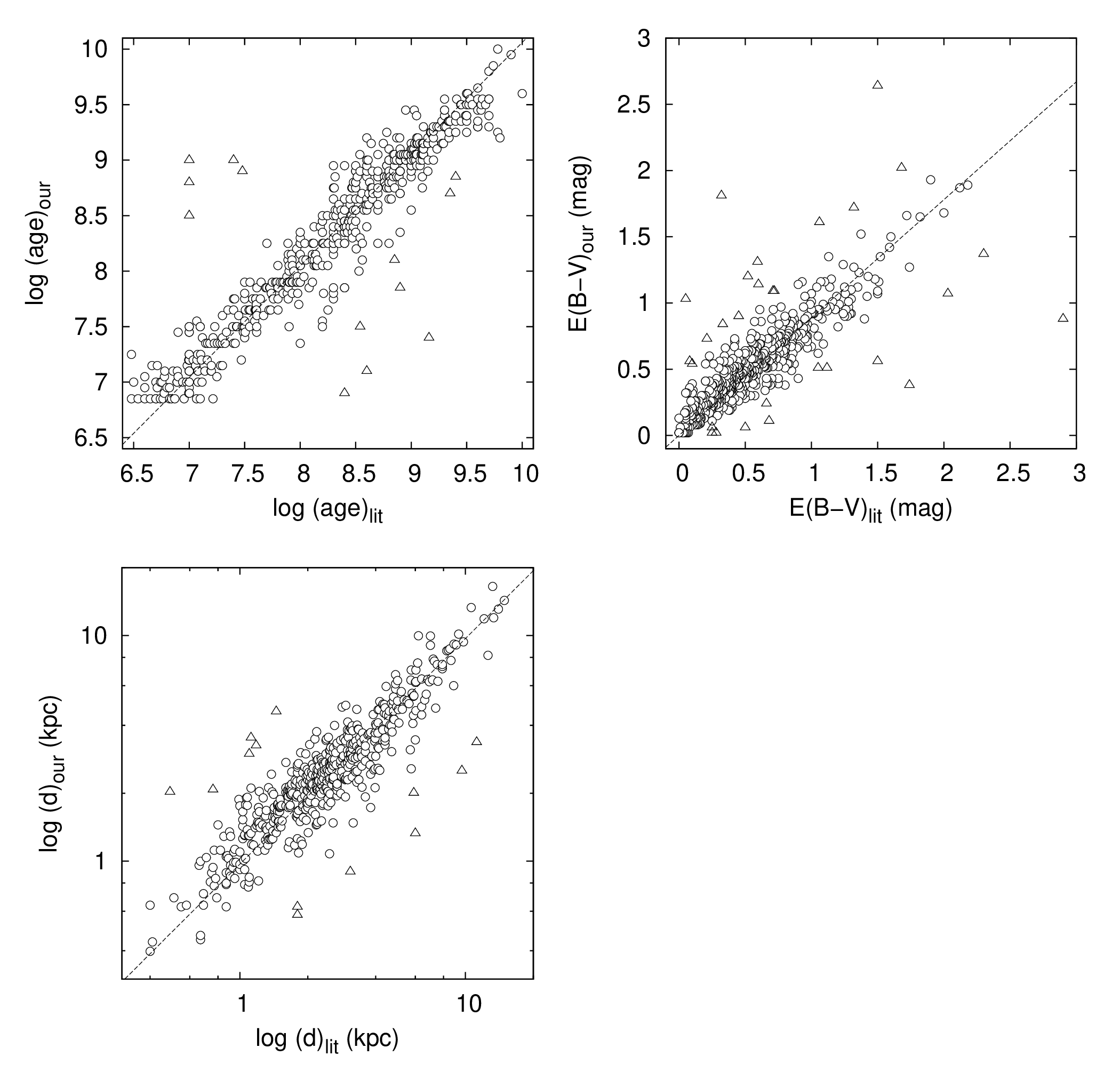}
\end{center}
\FigCap{a) Open clusters ages derived by us as a function of the
literature data. The best--fitting linear relation of a the form $y=ax$
is marked with a dashed line. \textbf{Triangles} were rejected during fitting a
linear trend, see text for details. b) The same as panel a but for the
interstellar reddening $E(B-V)$. c) The same as panel a but for the
distance from the Sun $d$.} 
\end{figure}

To calculate the interstellar reddening $E(B-V)$, we
adopted the interstellar extinction law by Schlegel et al.~(1998) \textbf{and Bessell (2005)}
$\mathitbf{E(J - K_{S}) =}~\mathbf{0.972~E(J - K)}$ \textbf{assuming} $\mathitbf{E(J-K_{S}) / E(B-V) =}~\mathbf{0.520 \pm 0.045}$.
In Fig.~2b we plot the comparison of the obtained colour excess $E(B-V)$ with the
literature data for 543 clusters. Twenty nine outlying clusters, which
were found to be located too far from the main trend, were not taken
into account during fitting procedure. We obtained a linear relation:

\begin{equation}
\centering
E(B-V)_{our} = (\mathbf{0.89}\pm0.01)\:E(B-V)_{lit}
\end{equation}
with the correlation coefficient equal to \textbf{0.91}. Our determinations were
found to be somewhat lowered comparing to the published ones. This
suggests slightly different ratio $E(J-K_{S}) / E(B-V) = \mathbf{0.463}$.

We  calculated distances to the clusters assuming relations
$E(J-K_{S}) / E(B-V) = \mathbf{0.520}$ and $A_{J}/A_{V} = 0.276$
and under the assumption of the total-to-selective absorption ratio of
$R = 3.1$ (Cardelli et al.~ 1989). In  Fig.~2c the distances, $d$, are
plotted against published data for 532 objects. Almost all points are
located along a diagonal line corresponding to a perfect match. After
removing thirteen outlying objects, we obtained a linear relation:

\begin{equation} \centering d_{our} = (0.97\pm0.09)\:d_{lit}
\end{equation} 
resulting in the correlation coefficient of 0.93. This
result allows us to find our determinations of $d$ reliable. It is worth
noticing that adopting the relation $E(J-K_{S}) / E(B-V) = \mathbf{0.463}$ would
change $d$ only by \textbf{2.8}{\%} -- a value two times smaller than typical
errors based on the distance modulus.


\section{Discussion of statistical relations between parameters}

\subsection{Reddening, age, radial structure and distribution in the Galactic Plane}

Structural and physical parameters determined for statistically
significant number of open clusters allowed us to check mutual relations
between individual parameters. Fig.~3a) presents a relation between the
interstellar reddening ($E(B-V)$) and a distance from the Galactic Plane
($Z$). The presented relation indicates that for objects located within
$Z = \pm 0.1$ kpc the average reddening is equal to \textbf{0.68} mag and
decreases with increasing  $|Z|$. For clusters with $|Z|$ between 0.1 and
0.4 kpc the mean $E(B-V)$ is equal to \textbf{0.51} mag and drops to \textbf{0.34} mag for
$|Z| > 0.4$ kpc. We performed similar statistical analysis for a
distribution of $E(B-V)$ along cluster distances from the Galactic
Centre, $R_{GC}$ (Fig.~3b). The mean $E(B-V)$ for open clusters located
near the Sun (8.3 kpc) and far beyond the Galactic orbit of the Sun is
\textbf{0.74} and \textbf{0.53} mag, respectively. Both trends \textbf{were noticed by Janes \& Phelps (1994)}.

\begin{figure}[htb]
\begin{center}
\includegraphics[width=1.0\textwidth]{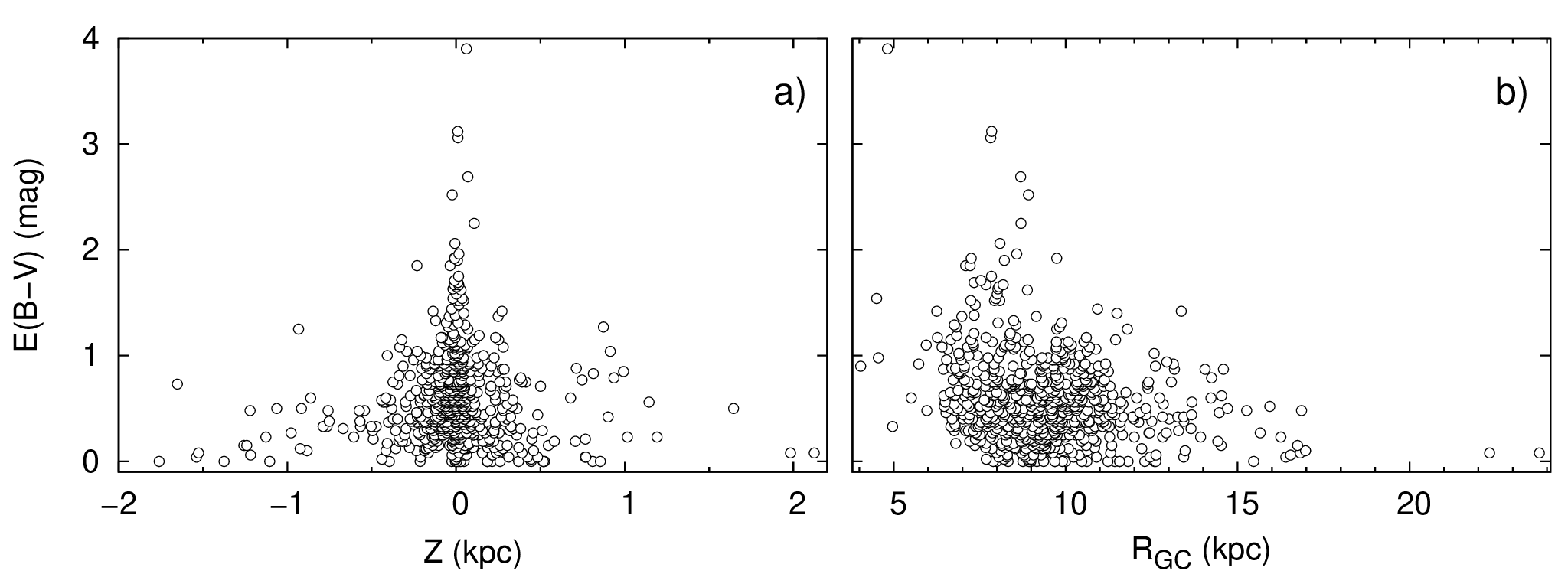}
\end{center}
\FigCap{\emph{Left}: The relation between the reddening $E(B-V)$ and
distance from the Galactic Plane $Z$ assuming
$Z_{\odot} = -33$pc. \emph{Right:} Relations between the reddening
$E(B-V)$ and distance from the Galactic Centre $R_{GC}$, assuming
$R_{GC\odot} = 8.3$kpc.}
\end{figure}

Linear limiting radii \textit{vs.} linear core radii are plotted in
Fig.~4. We fitted a linear trend which resulted in an equation:

\begin{equation}
\centering
R_{lim} = (\mathbf{6.12}\pm0.12)\:R_{core} + (\mathbf{1.28}\pm0.14)\:
\end{equation}
where both $R_{lim}$ and $R_{core}$ are in pc. The correlation
coefficient was found to be \textbf{0.88}. Similar relations were obtained in
studies from near IR data: Nilakshi et al.~(2002) studied 38 open
clusters and obtained the relation $R_{lim} = 6\:R_{core}$. It is worth noticing that
for 81{\%} of open clusters studied in this paper, the limiting and core
radii were smaller than 10 and 1.5 pc, respectively.

\begin{figure}[htb]
\begin{center}
\includegraphics[width=0.5\textwidth]{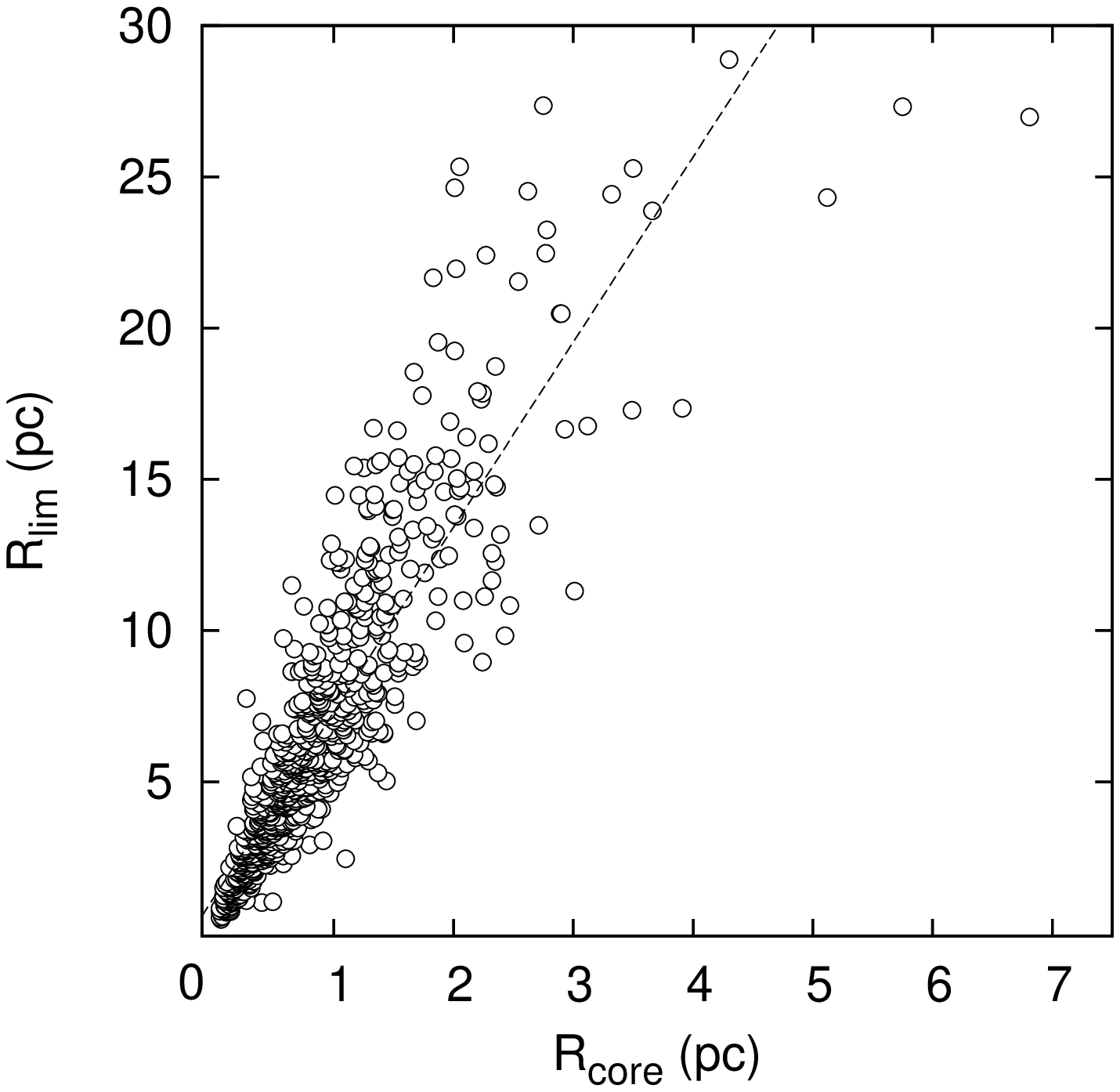}
\end{center}
\FigCap{The relation between the linear limiting and core radii for clusters of the sample.}
\end{figure}

While moving away from the Galactic Plane\textbf{,} we observed a higher fraction
of open clusters with greater limiting and core radii (Fig.~5a and b).
The similar conclusion was reached by Tadross et al.~(2002) and
Schilbach et al.~(2006). \textbf{For clusters younger than 500 Myr we obtained
a scale height perpendicular to the Galactic
Plane of 71 pc, for oldest 238 pc (Fig 5c)}. We notice that the
average age of clusters increases with a distance from the Galactic
Centre \textbf{(Fig.~5d)}. After removing nine outlying objects\textbf{,} the average
$R_{GC}$ of about 9.0 kpc and for the old ones of 10.0 kpc,
assuming the distance from the Galactic Centre to the Sun
$R_{GC\odot} = 8.3$ kpc. We binned the distribution of individual
sizes of clusters into 0.3 log ($age$) bins (Fig.~6) and noticed that
the average linear sizes increase with age, but we observed \textbf{also} strong
selection effect (see Sect. 4.2) which affects the relation between
cluster sizes and their ages.

\begin{figure}[htb]
\begin{center}
\includegraphics[width=1.0\textwidth]{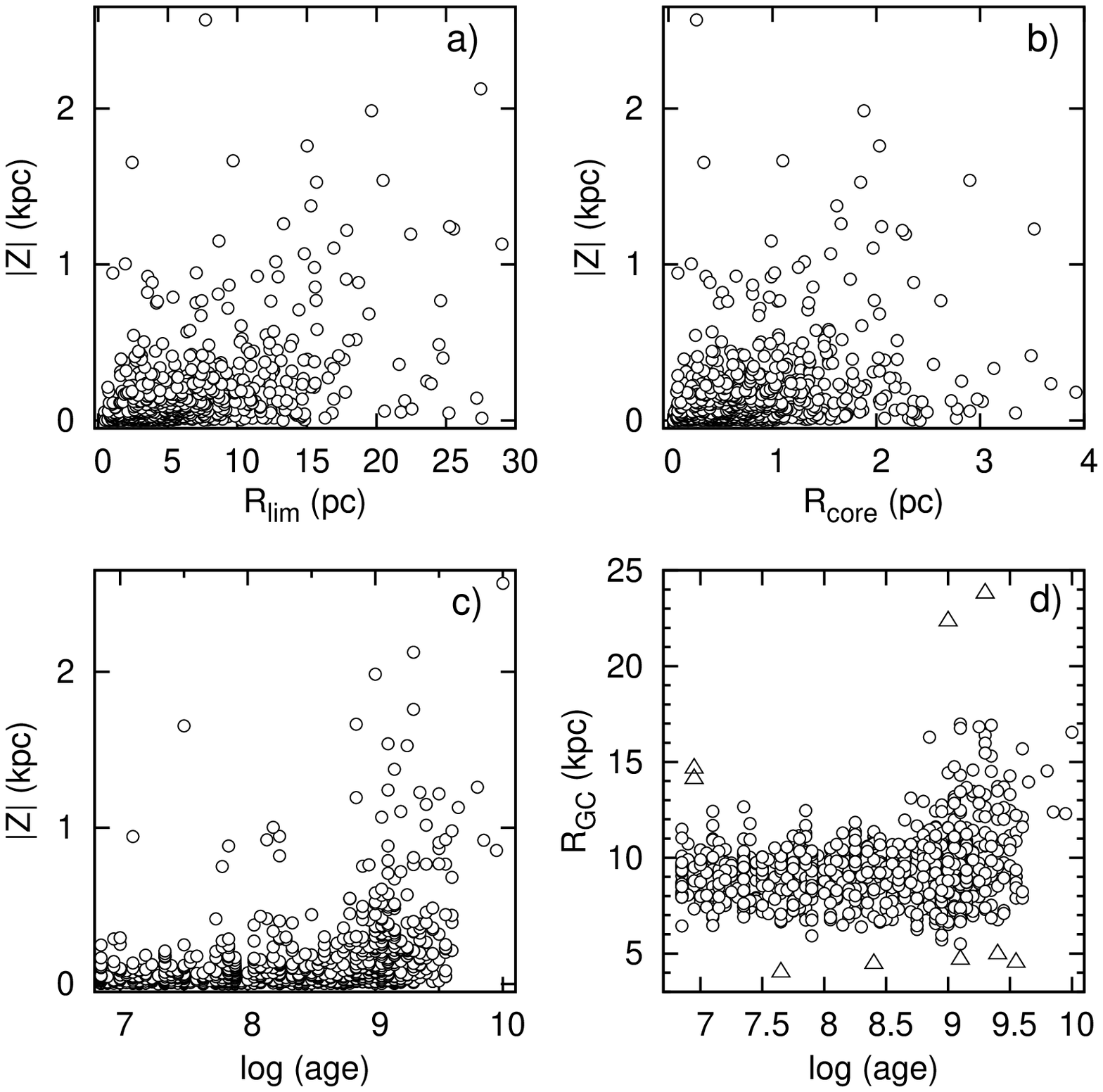}
\end{center}
\FigCap{a) Relation between the limiting radius in parsecs and distance
from the Galactic Plane $Z$, assuming $Z_{\odot} = -33$ pc for the Sun.
b) The same as panel a but for core radii. c) Relation between the age
of open clusters and distance from the Galactic Plane $Z$. d) Relation
between the age of open clusters and distance from the Galactic Centre
$R_{GC}$.}
\end{figure}

\begin{figure}[htb]
\begin{center}
\includegraphics[width=1.0\textwidth]{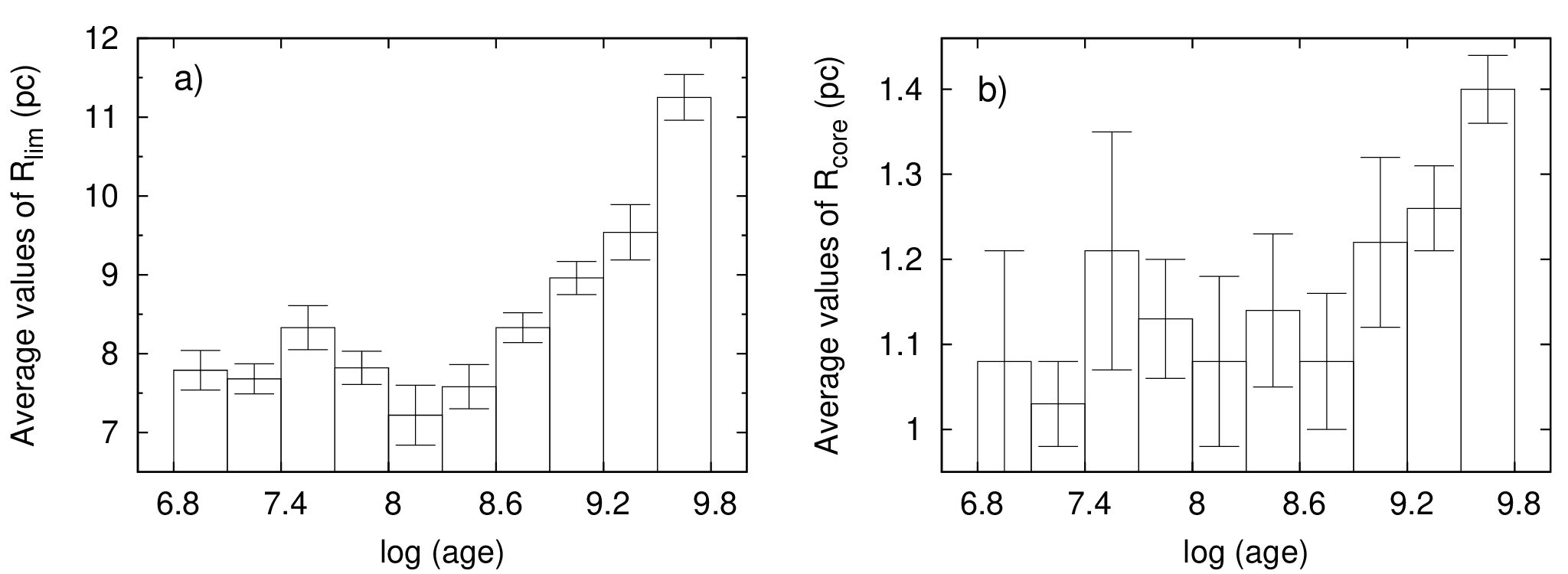}
\end{center}
\FigCap{\emph{Left:} relation between the limiting radius in parsecs of
open clusters and their ages. Results were presented as the average
values of sizes in the particular range of the age. \emph{Right:}
relation between the core of open clusters and their ages. While
constructing both diagrams we removed three clusters with log ($age$)
above 9.8. We used Sturgesa's method (1926) to produce both histograms.}
\end{figure}

Geometrical and physical characteristics of statistically significant
sample of open clusters also allowed us to check some large spatial
scale properties of the system of open clusters in the Galaxy. \textbf{Our
results presented in Fig.~7 reveal two regions at
the Galactic longitude with a significantly diminished number
of clusters what confirms results reported by Tadross et al.
(2002) and Froebrich et. al (2010). A first region is located near 
the Galactic Centre, i.e. $\mathbf{60^{\circ}}$ away from it}. This is probably
an effect of observational selection because at $0^{\circ}$
we observe in the direction of the Galactic Centre where high stellar
background and gas density may effectively prevent from detecting open
clusters. A second region with a number of cluster smaller than average
is seen at $140\,^{\circ} < l < 200\,^{\circ}$. The same effect is also observed in
the distribution of stars shown by Benjamin (2008), who discovered a gap
in the Perseus Spiral Arm. Froebrich et al.~(2010) noted that in
all-sky extinction maps by Rowles and Froebrich (2009) there is no
indication of the existence of such high-extinction molecular clouds in
this region. Additionally, we notice that a \textbf{second gap in the cluster distribution}
is more pronounced for younger
objects. Oldest clusters have orbited the Galactic Centre more times
than younger ones so their distribution in the Galactic Plane is less
related to the region where they have been born. Thus, this lack might
be connected with a local gap in the Perseus Spiral Arm.

\begin{figure}[htb] 
\begin{center}
\includegraphics[width=0.6\textwidth]{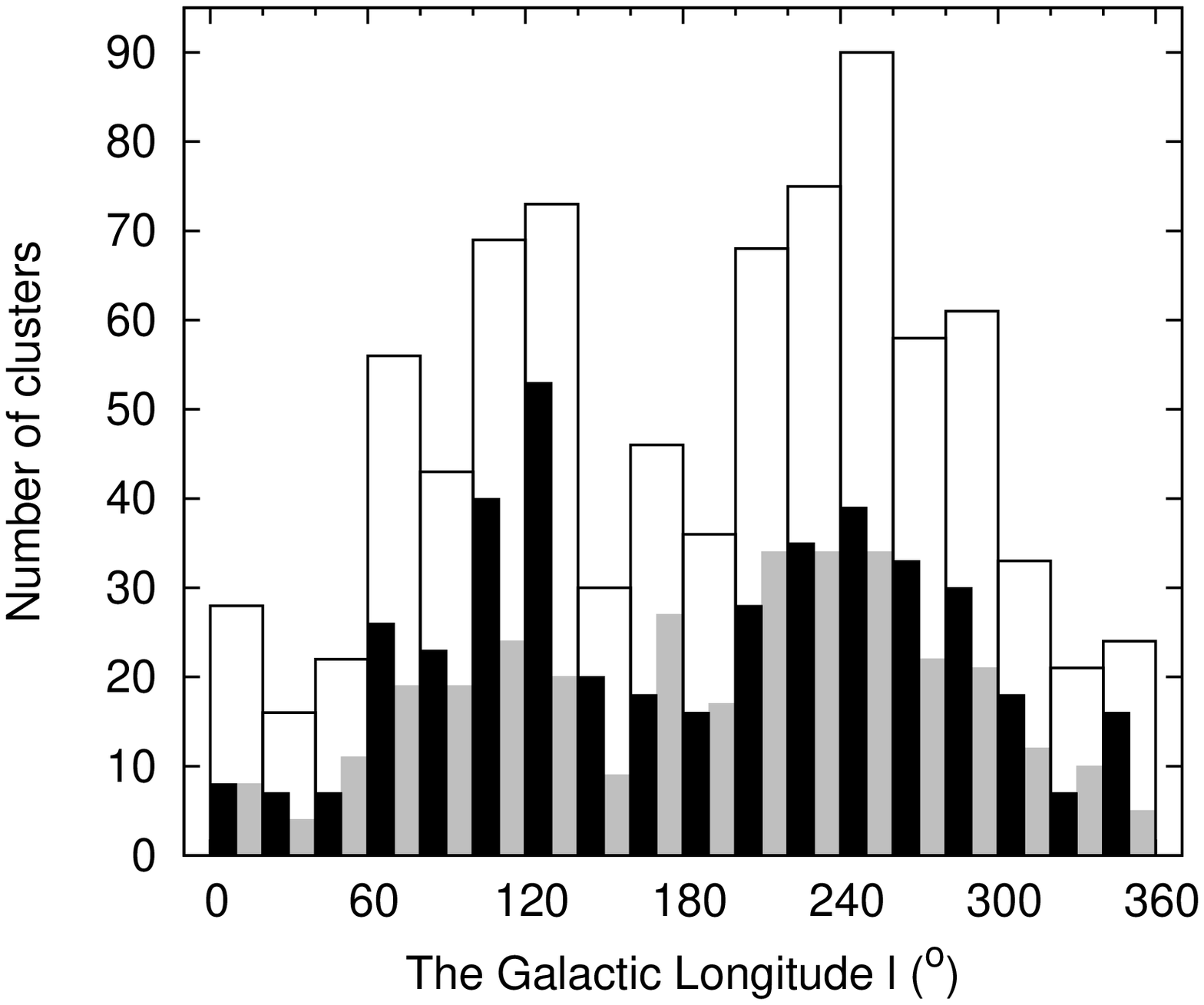} 
\end{center} 
\FigCap{Histogram showing the distribution of the clusters of our sample along
the Galactic Plane. Black, grey, and white rectangles represent young
($\log (age) < 8.7$), old ($\log (age) > 8.7$), and all open clusters,
respectively.}
\end{figure}

The concentration parameter is in a range between 0.8--1.1 for 94{\%} of
open clusters from our sample. We did not detect any relation between
$c$ and $|Z|$ or $R_{GC}$.

\begin{figure}[htb]
\begin{center}
\includegraphics[width=1.0\textwidth]{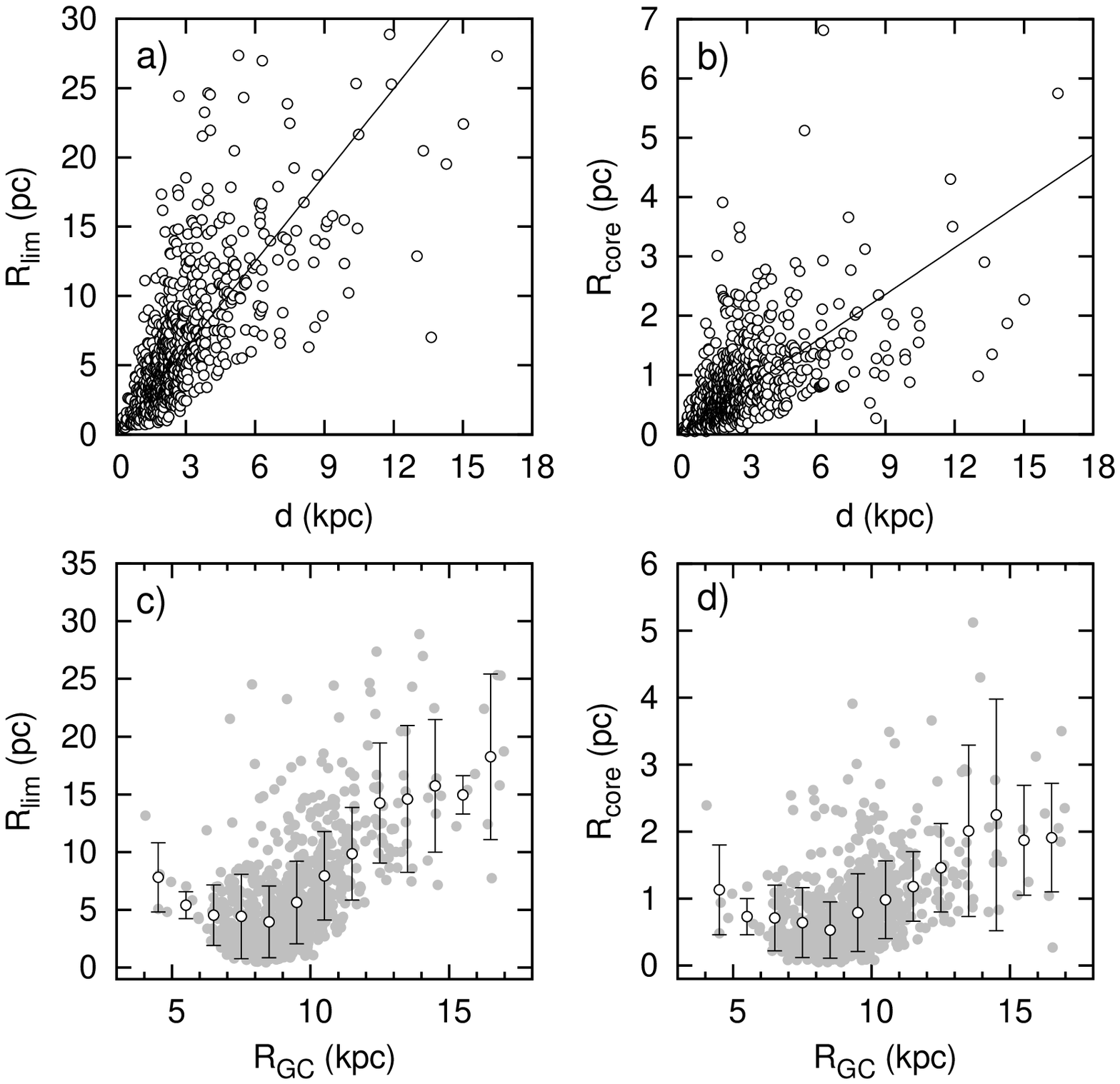}
\end{center}
\FigCap{a) Relation between the limiting radius and distance from the
Sun. b) The same but for core radii. The black lines indicate the
correlation as given in equations \textbf{9} and \textbf{10}. c) Relation
between the limiting radii and the distance from the Galactic Centre
$R_{GC}$. The Sun is located at $R_{GC\odot} = 8.3$ kpc. Gray points
represent individual measurements while black ones mean values in
succeeding bins. The bin size is 1 kpc. d) The same as c but for core
radii.}
\end{figure}

\subsection{Selection effects and biases}

The \textbf{first} main selection effect has its source in the  catalog by Dias et al.~
(2002),  from which the list of open
clusters was taken. It is a
compilation of many different catalogs what makes our sample  
inhomogeneous. \textbf{Second is the 2MASS catalog. The maximum range of
limiting magnitude of this survey is about 16.5 mag in the $J$ band.
Moreover the resolution is relatively low, what makes it difficult to study
small open clusters in high density fields. Another problem is} the
extinction along the line of sight. \textbf{Other bias may be} caused
by extracting sources with a radius of
34 arcmin around clusters. The selection radius might introduce a bias
in the cluster radius determination, in particular for larger clusters.
But this is not a significant problem because the median limiting radius
of the all clusters in our sample is 7 arcmin.

Selection effects make \textbf{the determination of the real limiting size of
clusters impossible}, so there is \textbf{a} possibility that all open clusters are bigger than
our determinations, and our $r_{lim}$ is a lower limit for cluster size.

Finally, we found important selection bias in our sample. In Fig.~8a and
b we show that cluster sizes increase with distance from the Sun. For
all clusters we obtained relations:

\begin{equation}
\centering
R_{lim}\:\mathbf{(pc)}\: = \mathbf{(2.080\pm0.038)}\:\:d\:\mathbf{(kpc)}
\end{equation}

\begin{equation}
\centering
R_{core}\:\mathbf{(pc)}\:= \mathbf{(0.262\pm0.007)}\:\:d\:\mathbf{(kpc)}
\end{equation}
with the correlation coefficient \textbf{0.67} and \textbf{0.47}, respectively. The same
problem was encountered by Froebrich et al.~(2010). This selection
effect causes  $R_{lim}$ and $R_{core}$  to increase with 
distance from the Galactic Centre, what is illustrated in Fig.~8c and d,
respectively. Similar trend has already been observed by Lyng\aa~(1982),
van den Bergh et al.~(1991), Tadross et al.~(2002) and Bonatto and Bica
(2010), but we argue that there is a strong selection effect in this
trend. When  observing distant open clusters we fail
to detect smaller ones, so all samples must exhibit a lack of
distant, compact open clusters. Hence our study of the evolution of
cluster sizes (in particular the limiting size) with age and position in
the Galaxy are encumbered with selection effects.


\section{Conclusions}  

The near--infrared $JHK_{S}$ photometric data
from the 2MASS \emph{Point Source Catalogue} (Skrutskie et al.~ 2006) was
used to determine structural and physical parameters of a large sample
of known open clusters. The structural properties have been determined
for the complete sample of 849 open clusters, while the age and distance
were obtained for 754 of them. 

We showed that open clusters studied in near-infrared seem to be larger
than in optical bands. The same property was noticed by Sharma et al.~
(2006). Near IR data allow us to detect faint cluster members located
far from the cluster core and are not limited by the telescope's field
of view.

We obtained a different relation of $E(J - K_{S})/ E(B - V)$ than the
one given by Schlegel et al.~(1998) who obtained $E(J - K_{S})/ E(B - V)
= \mathbf{0.520}$. Our results give $E(J - K_{S})/ E(B - V) = \mathbf{0.463}$. This
difference translates into the error of the distance determinations of
only \textbf{2.8}{\%}. We noticed that when moving outward from the Galactic
Plane and the Galactic Centre the reddening of clusters statistically
decreases. The mean $E(B - V)$ for $|Z| < 0.1$ kpc is \textbf{0.68} mag, between
0.1 and 0.4 kpc is equal to \textbf{0.51} mag and for $|Z| > 0.4$ kpc is only
\textbf{0.34} mag. We showed the similar trend for $R_{GC}$. For clusters located
near the Sun's orbit and further than 8.3 kpc from the Galactic Centre,
the mean $E(B - V)$ is \textbf{0.74} and \textbf{0.53} mag, respectively. This may be
implied by the fact that the dust and gas density decreases, too.

For the majority of studied clusters (81{\%}) the limiting radii was
found to be smaller than 10 pc and the core radii smaller than 1.5 pc.
Statistically the limiting radius is about 6--7 times larger than the
core radius. We also noticed that the average cluster sizes appear to
grow with a distance from the Galactic Centre and Galactic Plane. This
effect has already been observed by Lyng\aa~(1982), van den Bergh et al.~
(1991), Tadross et al.~(2002), Schilbach et al.~(2006) and Bonatto and
Bica (2010). Moreover, we noticed that the average age of clusters
increases with a distance from the Galactic Centre and the Galactic
Plane what was shown by Janes and Phelps (1994), Friel (1995), Tadross et
al.~(2002) and Froebrich et al.~(2010).

Open clusters, younger than 500 Myr, have a scale height of 71 pc, while
older clusters 238 pc. These results are similar to Janes and Phelps
(1994) who derived scale heights of 55 pc and 375 pc for young and old
clusters, respectively. We found that a number of older clusters
increases with the distance from the Galactic Centre. For younger
clusters we obtained an average $R_{GC} = 9.0$ kpc and for the older
ones -- $10.0$ kpc. This seems to confirm values of $8.9$ and $9.4$ kpc
for younger and older clusters, respectively, obtained by Froebrich et
al.~(2010) and by Tadross et al.~(2002) who found that younger clusters
are concentrated at $R_{GC} < 9.5$ kpc. This shows a difference between
younger and older clusters in their location. We also notice that older
clusters seem to be larger than the younger ones, what was shown by
Lyng\aa~(1982) and Tadross et al.~(2002).

Moreover we found important selection effects in our sample. Cluster
sizes grow with distance from the Sun, the same problem was encountered
in Froebrich et al.~(2010).

We observed a smaller number of clusters at the Galactic longitude range
of $140\,^{\circ} < l < 200\,^{\circ}$ what is in agreement with Tadross
et al.~(2002) and Froebrich et al.~(2010). This may reflect the
structure of the Galaxy in this direction where there is a gap in the
Perseus Arm (Benjamin 2008).

\Acknow{This research is supported by "Stypendia dla doktorant\'ow
2008/2009 -- ZPORR" SPS.IV-3040-UE/204/2009. This publication makes use
of data products from the Two Micron All Sky Survey, which is a joint
project of the University of Massachusetts and the Infrared Processing
and Analysis Center/California Institute of Technology, funded by the
National Aeronautics and Space Administration and the National Science
Foundation.}



\newpage
\appendix
\section{Parameters of studied open clusters}

\footnotetext{Full data tables available at http://www.astri.uni.torun.pl/\textasciitilde gm/OCS/2mass.html}

\MakeTable{lcccccccc}{13cm}{The list of the studied open clusters the new equatorial and galactic coordinates for epoch J2000.0, $\alpha$ and $\delta$ -- the equatorial coordinates, $l$ and $b$ -- the Galactic Longitude and Latitude, $r_{lim}$ -- the limiting radius (angular), $r_{core}$ -- the core radius (angular), $R_{lim}$ -- the limiting radius (linear), $R_{core}$ -- the core radius (linear), $c$ -- the concentration parameter.}
{
\hline
\hline
Star cluster 	&$\alpha$~~~~~~~~~~~~~~~$\delta$&	$l$	&	$b$	&$r_{lim}$	&$r_{core}$&$R_{lim}$	&$R_{core}$	&$c$\\
		&$hhmmss\pm ddmmss$&$(\,^{\circ})$	&$(\,^{\circ})$	&(')			&(')&(pc)&(pc)&\\
\hline
Berkeley 58&000013+605619&116.7498&-1.3168&11.9$\pm$0.8&2.04$\pm$0.16&9.32$\pm$1.22&1.60$\pm$0.23&0.76\\
Stock 18&000136+643724&117.6224&2.2666&6.0$\pm$0.4&0.37$\pm$0.03&2.17$\pm$0.27&0.14$\pm$0.02&1.20\\
Berkeley 104&000328+633546&117.6289&1.2194&5.2$\pm$0.5&0.51$\pm$0.04&7.39$\pm$1.20&0.73$\pm$0.10&1.01\\
Czernik 1&000744+612823&117.7376&-0.9571&1.7$\pm$0.2&0.26$\pm$0.03&0.88$\pm$0.16&0.14$\pm$0.02&0.81\\
Berkeley 1&000946+602851&117.8173&-1.9764&7.3$\pm$0.6&0.94$\pm$0.07&7.38$\pm$1.07&0.95$\pm$0.13&0.89\\
King 13&001019+611031&117.9946&-1.3017&15.6$\pm$1.0&2.53$\pm$0.16&13.45$\pm$1.70&2.18$\pm$0.27&0.79\\
Juchert-Saloran 1&001619+595758&118.5456&-2.6062&8.7$\pm$0.6&1.09$\pm$0.07&9.80$\pm$1.33&1.23$\pm$0.16&0.90\\
Berkeley 60&001744+605616&118.8484&-1.6665&8.4$\pm$0.8&1.10$\pm$0.08&11.01$\pm$1.80&1.44$\pm$0.20&0.88\\
\hline
\hline
}


\MakeTable{lccccccc}{13cm}{The list of the studied open clusters: log ($age$) -- the logarithm of age, $m-M$ -- the distance modulus, $E(J - K_{\mathbf S})$ and $E(B - V)$ -- the reddening, $d$ -- the distance from the Sun, $Z$ -- a distance from the Galactic Plane, with assumption $Z$ for the Sun $Z_{\odot} = -33$ pc, $R_{GC}$ -- a distance from the Galactic Centre with assumption $R_{GC}$ for the Sun $R_{GC\odot} = 8.3$ kpc. An error for the age, m-M and reddening we assumed respectively \textbf {0.05}, 0.1 and 0.03 for more details please see the paper (Sect. 2.4.)}
{\hline\hline
Star cluster 	&Age&m-M&$E(J-K_{\mathbf S})$& $E(B-V)$&$d$&$Z$&$R_{GC}$\\
		& log ($age$)&(mag)&(mag)&(mag)&(kpc)&(kpc)&(kpc)\\
\hline
Berkeley 58&8.25&12.70&0.34&0.64&2.699$\pm$0.178&-0.062$\pm$0.004&9.815$\pm$0.077\\
Stock 18&8.10&11.08&0.37&0.69&1.252$\pm$0.083&0.050$\pm$0.003&8.949$\pm$0.038\\
Berkeley 104&8.85&13.83&0.23&0.43&4.925$\pm$0.310&0.105$\pm$0.007&11.448$\pm$0.132\\
Czernik 1&7.25&11.79&0.32&0.60&1.801$\pm$0.118&-0.030$\pm$0.002&9.276$\pm$0.054\\
Berkeley 1&9.00&13.16&0.29&0.54&3.461$\pm$0.223&-0.119$\pm$0.008&10.377$\pm$0.099\\
King 13&8.90&12.81&0.28&0.52&2.968$\pm$0.190&-0.067$\pm$0.004&10.041$\pm$0.086\\
Juchert-Saloran 1&9.05&13.64&0.43&0.80&3.894$\pm$0.266&-0.177$\pm$0.012&10.721$\pm$0.120\\
Berkeley 60&8.25&14.04&0.48&0.90&4.512$\pm$0.315&-0.131$\pm$0.009&11.198$\pm$0.142\\
\hline\hline}

\newpage
\section{Colour--magnitude diagrams}

\begin{figure}[htb]
\begin{center}
\includegraphics[width=0.89\textwidth]{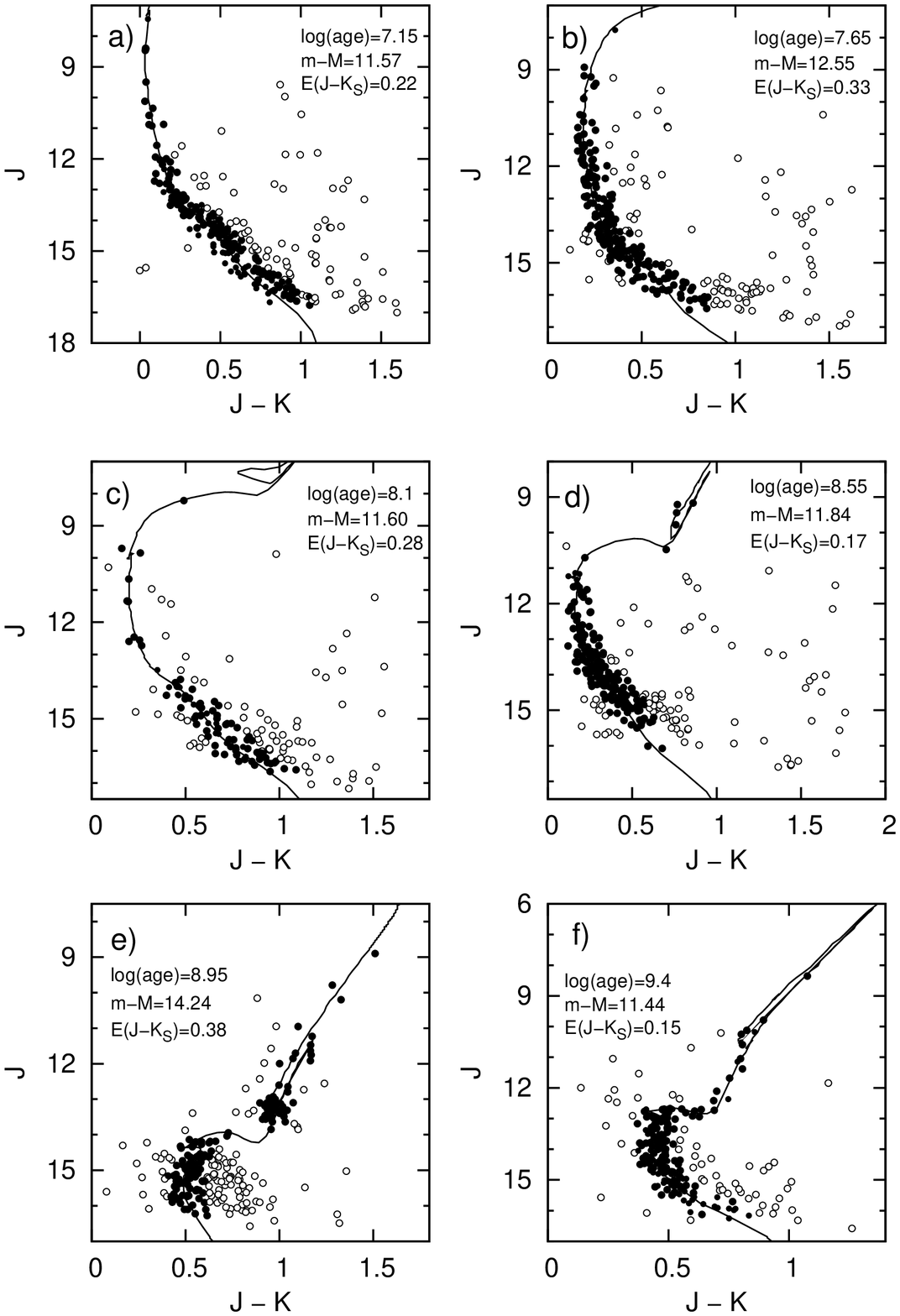}
\end{center}
\FigCap{Colour--magnitude diagrams after the cleaning procedure for six
different open clusters. Filled circles represent stars used for fitting isochrones while
open circles -- stars manually removed after visual inspection.
The best-fit isochrones are drawn
with solid lines. a) NGC 2645 b) NGC 663 c) Markarian 50 d) Melotte 105 e)
IC 166 f) NGC 6253. For more information about open clusters please see
Table 1 and 2.}
\end{figure}

\end{document}